\documentclass[12pt]{article}

\usepackage[latin1]{inputenc}
\usepackage{amsmath}
\usepackage{amsfonts}
\usepackage{amssymb,MnSymbol,slashed}
\usepackage{cite}
\usepackage{amsxtra}
\usepackage[margin=3cm]{geometry}
\usepackage{color}
\usepackage{hyperref}
\hypersetup{linktocpage=true}
\usepackage{graphicx}
\usepackage{placeins}
\usepackage{upgreek} 
\usepackage{mathrsfs}
\usepackage{accents}
\usepackage{multirow} 
\usepackage{tikz}

\numberwithin{equation}{section}
\newcommand{\beq}{\begin{equation}}
\newcommand{\eeq}{\end{equation}}
\def\be {\begin{equation}}
\def\ee {\end{equation}}
\def\bs#1\es{\begin{split}#1\end{split}}
\def\ba#1\ea{\begin{align}#1\end{align}}
\def\baed#1\eaed{\begin{aligned}#1\end{aligned}}
\def\bged#1\eged{\begin{gathered}#1\end{gathered}}
\def\bea{\begin{eqnarray}}
\def\eea{\end{eqnarray}}

\def\nn{\nonumber}

\def\a{\alpha}

\def\d{\delta}

\def\e{\epsilon}

\def\f{\phi}

\def\G{\Gamma}
\def\h{\eta}

\def\m{\mu}
\def\n{\nu}

\def\t{\tau}

\def\z{\zeta}

\def\bls{\bigg [}
\def\brs{\bigg ]}

\def\cL{\mathcal{L}}

\def\cO{{{\mathcal O}}}
 
\def\cN{\mathcal{N}}
\def\cV{\mathcal{V}}

\def\bZ{\mathbb{Z}}
\def\bR{\mathbb{R}}

\def\pr{\prime}

\def\pa{\partial}
\def\na{\nabla}
\def\fr{\frac}

\def\we{\wedge}
\def\ra{\rightarrow}

\def\lra{\leftrightarrow}

\newcommand{\wh}[1]{ {\hat{#1}}{} }
\newcommand{\til}[1]{ {\tilde{#1}} }

\let\foo\bar 
\renewcommand{\bar}[1]{ {\foo{  #1} }{} }

\newlength{\dhatheight}

\def\tbzero{{\text{\tiny{(0)}}}}

\def\tbthree{{\text{\tiny{(3)}}}}
\def\alpr{{\a^\pr}}

\begin{document}

\baselineskip=16pt
\setlength{\parskip}{6pt}

\begin{titlepage}
\begin{flushright}
\parbox[t]{1.4in}{
\flushright IPhT-T15/107}
{\flushright MPP-2015-127}
\end{flushright}

\begin{center}

\vspace*{1.7cm}

{\LARGE \bf  F-theory at order $\alpha'^{\,3}$}

\vskip 1.6cm

\renewcommand{\thefootnote}{}

\begin{center}
 \normalsize 
 Ruben Minasian$^a$, Tom G. Pugh$^b$ and Raffaele Savelli$^a$
\end{center}
\vskip 0.1cm
$^a$ {\sl Institut de physique th\'eorique, Universit\'e Paris Saclay, CEA, CNRS, F-91191 Gif-sur-Yvette,
France}

$^b$ {\sl Max Planck Institute for Physics,
F\"ohringer Ring 6, 80805 Munich, Germany}
\end{center}

\vskip 1.5cm
\renewcommand{\thefootnote}{\arabic{footnote}}

\begin{center} {\bf ABSTRACT } \end{center}
We study the effective physics of F-theory at order $\a'^{\,3}$ in derivative expansion.   We show that the ten-dimensional type IIB eight-derivative couplings involving the graviton and the axio-dilaton naturally descend from pure gravity in twelve dimensions. Upon compactification on elliptically fibered Calabi-Yau fourfolds, the non-trivial vacuum profile for the axio-dilaton leads to a new, genuinely $\cN=1$, $\a'^{\,3}$ correction to the four-dimensional effective action.

\end{titlepage}

\newpage
\noindent\rule{\textwidth}{.1pt}		
\tableofcontents
\vspace{20pt}
\noindent\rule{\textwidth}{.1pt}

\setcounter{page}{1}
\setlength{\parskip}{9pt}

\section{Introduction and outlook}

F-theory \cite{Vafa:1996xn} provides an elegant framework to combine all key ingredients for realistic models of particle physics. At the same time, type IIB string theory with D7-branes/O7-planes, of which F-theory is the strongly coupled formulation, is one of the corners of the string landscape where the process of lifting vacuum degeneracies is best understood (see \cite{Denef:2008wq,Maharana:2012tu} for comprehensive reviews).

As opposed to other types of moduli, the K\"ahler moduli of type IIB compactifications are stabilized by quantum effects. Within the Large Volume Scenario \cite{Balasubramanian:2005zx}, in particular, besides euclidean D-brane instantons, a crucial role is played by the leading $\a'$ corrections to the effective four-dimensional (4d) K\"ahler potential. Due to the (extended) no-scale structure of these models, it turns out that the main impact on effective masses and couplings typically comes from a certain $\a'^{\,3}$ correction\footnote{Corrections to the K\"ahler potential at order $\a'^{\,2}$ have been computed in toroidal orbifold models \cite{Berg:2005ja}, and their general behavior studied in the blown-up phase. Some models are robust against inclusion of such corrections, due to certain cancellations happening at the level of the scalar potential \cite{vonGersdorff:2005bf,Berg:2005yu,Cicoli:2007xp}. See \cite{Berg:2007wt} for a review.}, which originates from reducing known ten-dimensional (10d) $R^4$ couplings on Calabi-Yau (CY) threefolds\footnote{This analysis has been extended lately to compactifications on $SU(3)$ structure geometries \cite{Grana:2014vva}.} and performing the Weyl rescaling needed to bring the 4d action into the Einstein frame \cite{Antoniadis:1997eg,Becker:2002nn,Antoniadis:2003sw}. For a CY threefold $X_3$ of Euler number $\chi(X_3)$ and classical volume $\cV_3$ in units of $2\pi \sqrt{\alpha'}$, this has the general form
\ba
\label{BBHL}
K&=-2\log\left(\cV_3+\frac{\xi}{g_s^{3/2}}\right) \, ,&
\xi&=-\frac{\zeta(3)}{32\pi^3}\,\chi(X_3)\,,
\ea
where we have only displayed its tree-level part in string perturbation theory, which is the dominant one if the string coupling is stabilized at values $g_s\ll 1$. An important feature of the correction \eqref{BBHL} is that it originates from the underlying $\cN=2$ sector of closed strings propagating on CY threefolds, and hence it does not capture the effects of D7-branes and O7-planes, which are responsible for halving the amount of preserved supersymmetry. One therefore expects additional contributions to \eqref{BBHL}, which are purely $\cN=1$ corrections and which are sensitive to the 7-brane content of the theory. None of these corrections are known yet\footnote{Recently, new $\alpha'^{\,3}$ corrections to the 4d scalar potential of type IIB orientifolds, without counterparts in the K\"ahler potential, have been indirectly inferred from the known 10d $R^4$ couplings  \cite{Ciupke:2015msa}.}, and one of the aims of the present paper is to investigate on possible modifications of \eqref{BBHL} within the framework of F-theory. 

We will show that, for very small values of the string coupling, the orientifold background affects this $\a'^{\,3}$ correction in a way that amounts to replacing
\bea\label{N=1BBHL}
\chi(X_3)&\longrightarrow&\chi(X_3)\;+\;2\int_{X_3}D_{\rm O7}^3\,,
\eea
where $D_{\rm O7}$ is the class Poincar\'e dual to the divisor wrapped by the O7-plane in $X_3$. At order-one string coupling, i.e. in F-theory, these two terms merge into a single correction, which however is non-topological (see section \ref{sec:4d}).

In this sense, our computation generalizes \cite{Antoniadis:1997eg} to $\cN=1$ compactifications. However, this analysis is not sufficient to make statements about the $\cN=1$ metric for the K\"ahler moduli, and must be supplemented, in this context, by a more complete reduction which includes the kinetic terms for the scalars. Corrections to the K\"ahler metric at one-loop level have been recently derived in \cite{Berg:2014ama}, through a two-point function computation of closed string vertex operators in toroidal orientifold models. Our analysis is complementary to \cite{Berg:2014ama}, and it would be very important, especially for phenomenological studies of string vacua, to extrapolate the results of \cite{Berg:2014ama} to the smooth CY regime and combine them with those presented here. 

Another caveat is that our study neglects the warp factor. Recently, warping effects were considered jointly with higher derivative corrections, in the context of M-theory reductions on CY fourfolds \cite{Grimm:2014xva,Grimm:2014efa}. Progress has also been made in identifying the warped K\"ahler potential of F-theory/type IIB compactifications \cite{Martucci:2014ska}. It would be of great interest to use these results to combine the $\alpha'$ effects discussed here with warping and infer the consequences for the 4d, $\cN=1$ effective action. These effects will give a separate contribution to the K\"ahler potential, dependent on the warp factor, which will represent a different structure and cannot cancel the correction we describe in this paper. Nevertheless, this extra contribution could be equally relevant when studying phenomenological effects. We hope to report on this analysis in a future publication.

Our main focus in this paper are $\alpha'^{\,3}$ corrections in F-theory, from which the result \eqref{N=1BBHL}, valid for type IIB CY orientifold vacua, arises upon taking a weak $g_s$ coupling limit. The 4d, $\cN=1$ effective action of F-theory compactifications on CY fourfolds was first discussed at lowest order in $\alpha'$ in \cite{Grimm:2010ks}. Later, there have been a number of efforts in deriving $\alpha'$ corrections to these theories \cite{Grimm:2012rg,GarciaEtxebarria:2012zm,Grimm:2013gma,Grimm:2013bha,Junghans:2014zla}. In  most of these studies\footnote{In \cite{GarciaEtxebarria:2012zm} Heterotic/F-theory duality was used instead to derive all perturbative $\alpha'$ corrections to a specific 4d, $\cN=2$ F-theory compactification. In \cite{Junghans:2014zla}, various terms in D-brane/O-plane actions were analyzed to extract $\alpha'^{\,2}$ corrections to the $\cN=1$ K\"ahler potential.} the effective physics of F-theory was examined by exploiting the duality with M-theory: One first considers M-theory compactified on elliptically fibered CY fourfolds, and then sends the volume of the torus fiber to zero, which turns the $\cN=2$ effective action in three dimensions (3d) to an $\cN=1$ one in 4d. Under this zero-size limit (F-theory limit), the eleven-dimensional (11d) Planck length $l_M$ becomes infinitely smaller than the string length, and consequently not all $l_M$ corrections in 3d lift to $\alpha'$ corrections in 4d. The way the $l_M$ corrections in 3d are derived from reducing higher-derivative couplings in 11d supergravity does affect their behavior under the F-theory limit. While classical compactifications on elliptic fibrations were considered \cite{Grimm:2013gma,Grimm:2013bha}, one may perform ``quantum'' reductions which take into account non-zero Kaluza-Klein modes along the torus. This procedure was introduced for a constant torus  in \cite{Green:1997as} (see also \cite{Green:1997di} and \cite{Collinucci:2009nv}), and was proven to lead to corrections surviving the F-theory limit. This is indeed the 11d origin of the correction in \eqref{BBHL} \cite{Antoniadis:1997eg}. It is therefore natural to expect that the generalization of \eqref{BBHL} to include the effects of 7-branes would arise from extending the computation of \cite{Green:1997as} to the case of a non-trivially fibered torus.

In this paper, however, we adopt a different point of view on F-theory, which helps us bypassing eleven dimensions and thus the quantum reductions mentioned above. We are by no means proposing a more fundamental formulation of F-theory: Our approach essentially uses the original twelve-dimensional (12d) logic of Vafa's paper \cite{Vafa:1996xn} and pushes it to the eight-derivative level. We use 12d quantities to make manifest a larger symmetry group in 10d: The $SL(2,\mathbb{R})$ transformations of type IIB supergravity will be encoded in 12d diffeomorphisms. Nevertheless, we will not have a standard supergravity theory in 12d, as not all components of the 12d metric propagate\footnote{A somewhat similar point of view can be found in the recent works \cite{Choi:2014vya,Choi:2015gia}, which however do not go beyond two-derivative level.}: There will indeed be no lift of the 10d measure to 12d, and hence the overall size of the two additional directions, which always parametrize a torus, will not correspond to a physical degree of freedom.

The advantage of the 12d perspective is that it allows us to package all possible kinematical structures of 10d couplings of gravity and axio-dilaton into much simpler, purely gravitational 12d couplings. The easiest instance of this arises at two-derivative level, whereby both the 10d Einstein-Hilbert term and the kinetic term for the axio-dilaton can be seen to descend just from a 12d Einstein-Hilbert term, upon Kaluza-Klein reduction on a torus. In this paper we show that a similar phenomenon happens also at the eight-derivative level, where the relevant 12d structure we find has the familiarly looking form
\be\label{HatR}
\cL^{(3)}=f_0(\tau,\bar\tau)(\hat t_8 \hat t_8+\tfrac{1}{96} \hat \epsilon_{12} \hat \epsilon_{12})\hat{R}^4\,,
\ee
which contains various kinds of Lorentz-invariant contractions of four 12d Riemann tensors $\hat{R}$ (see section \ref{sec:10d}). Here $\tau$ fully determines the torus part of the 12d metric, and $f_0$ is a specific real analytic function encoding the string dynamics. Upon Kaluza-Klein reduction on a torus of fixed size, the simple 12d coupling \eqref{HatR} gives rise to an $SL(2,\mathbb{R})$-invariant set of eight-derivative couplings containing 10d gravity and gradients of the axio-dilaton. The four-point part of these terms was computed in string theory at tree-level, using the pure spinor formalism \cite{Policastro:2006vt,Policastro:2008hg}, and our result is in perfect agreement with this. Beyond four-point, formula \eqref{HatR} contains a prediction on the flux-less sector of the type IIB lagrangian at order $\alpha'^{\,3}$. It would be very important to test \eqref{HatR} further, by computing the relevant five-point functions in string theory.

The disadvantage of the 12d framework is that it can only capture the kinematical structure of the 10d couplings, and ignores the dynamical factors multiplying them. However, supersymmetry and $SL(2,\mathbb{Z})$ invariance are powerful enough to allow just a single dynamical structure which is given by the $f_0(\tau,\bar\tau)$ present in \eqref{HatR} \cite{Pioline:1998mn}. In order to actually derive the $f_0$ factor, rather than infer it, one needs to use the 11d perspective and generalize the quantum reduction of \cite{Green:1997as}  to external gravitons polarized along the torus.\footnote{We thank P. Vanhove for discussions concerning this point.}

Now that we have established \eqref{HatR} as the 12d F-theory effective lagrangian at order $\alpha'^{\,3}$ with zero flux, just like we do with the analogous $l_M^6$ lagrangian in 11d supergravity, we can use it to derive $\alpha'$ corrections to the 4d effective action by classically reducing it on an elliptically fibered CY fourfold. As in M/F-theory duality, the complex structure of the CY
takes into account the 7-brane data of the vacuum (through the pinching of the fiber over some loci of the base). However, the K\"ahler structure of the CY cannot be treated the same way as in M-theory. Therefore, we argue that, in our 12d approach, the physics is captured by the behavior of \eqref{HatR} over a 10d slice of the 12d fibration, i.e. by the action
\be
S^{(3)}=\int_{\mathbb{R}^{1,3}\times B_3}\cL^{(3)}|_{\mathbb{R}^{1,3}\times B_3}*_{10}1\,,
\ee
where the internal part of the 10d slice is diffeomorphic to the base $B_3$ of the elliptic CY. Note that, if we consider smooth elliptic fibrations in Weierstrass form, this is well-defined, since every fiber comes with a marked point and there is a holomorphic embedding of the base in the total space of the fibration (zero-section)\footnote{For fibrations with no sections, one may consider the associated Jacobian fibration, which always has a  zero-section \cite{Braun:2014oya}.}.

Let us conclude this introductory section with a final remark. It is possible to extend the 12d approach to include couplings of the three-form flux of type IIB string theory, $G_3=F_3-\tau H_3$. However, the analysis is trickier, as more kinematical and dynamical structures can appear in 10d, each of which need not be individually $SL(2,\mathbb{R})$-invariant like in \eqref{HatR}. Moreover, while the $G_3$ flux will give a contribution to the effective action, it will not contribute to the 4d Weyl rescaling which we are looking at in this paper, to the order we are considering. To see this, we note that fluxes could only contribute to the Weyl rescaling through their substitution into higher-derivative terms in 10d. These are already suppressed by a factor of $\a'^{\,3}$ compared with the two-derivative part. If the fluxes $F_3, H_3$ are quantized as $(F_3, H_3) = \a' (F_3^{\rm{flux}}, H_3^{\rm{flux}})$, so that $(F_3^{\rm{flux}}, H_3^{\rm{flux}})$ integrate over 3-cycles to order-one integers (as is the case for fluxes fixed by higher-derivative terms through tadpole cancellation), substituting this result into the higher-derivative terms will give a contribution suppressed by more than three powers of $\a'$, which is beyond the order we care about here.   A detailed analysis of such fluxes, within both the 12d and the 11d approaches, will hopefully be presented in a forthcoming publication. 

The structure of the paper is as follows. In section \ref{sec:10d} we will discuss the promotion to 12d of the 10d eight-derivative couplings of type IIB involving gravity and axio-dilaton. Expanding this result, we will then demonstrate that our 12d proposal reproduces all known four-point string amplitudes with gravitons and axio-dilatons as external legs \cite{Policastro:2006vt,Policastro:2008hg}. In section \ref{sec:4d} we will consider non-trivial vacuum profiles for the axio-dilaton and reduce our 12d result to 4d on elliptically fibered CY fourfolds. In this way, we will determine the $\cN=1$ correction to the 4d Weyl rescaling. In appendix \ref{Conventions} we summarize the conventions we use in performing our computations, whereas in appendix \ref{HigherDerivativeActions} we comment on the effect of field redefinitions on the eight-derivative corrections and record certain intermediate results.

\section{Type IIB couplings from twelve dimensions}
\label{sec:10d}

One is used to thinking of F-theory either as type IIB string theory in non-trivial axio-dilaton backgrounds or as a certain decompactification limit of M-theory on elliptically fibered manifolds. In this section, we will explore this duality in order to derive the 10d eight-derivative couplings of type IIB, involving Riemann tensors and derivatives of the axio-dilaton.

\subsection{Preliminaries}\label{sec:2.1}

We will start by reviewing the two-derivative 10d effective action of type IIB theory. The part which depends on the metric $g_{mn}$ and the axio-dilaton $\t=C_0+ie^{-\phi}$ is of the form
\ba
S^\tbzero = \int ( R - 2 P_m \bar P^m ) *_{10} 1 \,, 
\label{RandP2}
\ea 
where the index $m=0,\ldots,9$ and $R$ is the Ricci scalar for the metric $g_{mn}$.   In this expression the kinetic term for $\t$ has been packaged in terms of $P_m$ in such a way that the $SL(2, \bR)/U(1)$ coset structure of the action is made manifest. The relationship between $P_m$ and $\t$ is given by
\ba
P_m &= \fr{i}{2 \tau_2 } \na_m \t \,  ,& 
\bar P_m & = - \fr{i}{2 \tau_2 } \na_m \bar \t \,, & 
\tau = \tau_1 + i \tau_2 \, , 
\ea
and covariant derivatives of $P_m$ can be formed as
\ba
D_m P_n &= \na_m P_n  - 2 i Q_m P_n \, , & 
D_m \bar P_m & = \na_m \bar P_n  + 2 i Q_m \bar P_n \, , &
Q_m =  - \fr1{2\t_2} \na_m \t_1  = \fr{ i}{2} ( P_m - \bar P_m) \, . 
\label{covDdef}
\ea
It is useful to note that these definitions imply that $D_{[m} P_{n]} = 0 $. This will be true as long as $\tau$ remains a globally well-defined quantity. However, if there are 7-brane sources for $\tau$, $D_{[m} P_{n]} $ will instead pick up additional delta function contributions.

We note that the action \eqref{RandP2} should be multiplied by a factor of $1/ ( (2 \pi)^7 \alpha'^{\,4})$. In this section we will mainly be concerned with the functional forms of the actions we discuss, and so we will neglect such factors. However, they will become relevant in section \ref{sec:4d}, and we will consistently include them there. 

One of the key statements made by F-theory is that the axio-dilaton $\t$ may be associated with the complex structure of an auxiliary torus. This may be seen in that, if we define a 12d metric given by $\wh g_{M N}$ where $M = 0, \ldots, 11$ and where 
\ba
\wh g_{M N} =\left(  \begin{array}{ccc}
g_{m n} &0 & 0 \\
0 & \fr{1}{\t_2} & \fr{\t_1}{\t_2}\\
0 & \fr{\t_1}{\t_2} & \fr{\t_1^2 + \t_2^2}{\t_2}
\end{array} \right) \, , 
\label{metricdecomp}
\ea
the action \eqref{RandP2} can be rewritten as
\ba
S^\tbzero = \int  \wh R *_{10} 1 \, ,
\label{R12act}
\ea
where $\wh R$ is the Ricci scalar built from the metric $\wh g_{MN}$. There is no 12d promotion of the 10d measure, as the internal torus is taken to have fixed (unphysical) volume. We note here that only derivatives with respect to 10d directions are allowed in our analysis, such that $ \pa_M  = \left(  \begin{array}{ccc}
\pa_m &0 & 0 \\
\end{array} \right) $. 

If we had not known the two-derivative $\t$-dependent terms in the 10d action, but we had known that it should be possible to rewrite the action in terms of $\wh g_{M N}$, then we could have used the above analysis to derive the action \eqref{RandP2}. We would begin with the Einstein-Hilbert term in 10d, promote this to the 12d Einstein-Hilbert term \eqref{R12act}, and then by expanding this expression in $\t$ we would derive \eqref{RandP2}.  

Alternatively, we could have taken the 11d Einstein-Hilbert term and we could have reduced it on a torus with metric:
\be
 \frac{\nu}{\rm{Im} \tau} \left(  \begin{array}{cc}
1 & \rm{Re} \tau\\
 \rm{Re} \tau& |\tau|^2
\end{array} \right) \, , 
\ee
where $\nu$ is the torus volume, which in this case is physical. Using the relation which links $l_M$ to $\alpha'$ through $\nu$, we would have then exactly obtained \eqref{RandP2} by taking the F-theory limit $\nu \rightarrow 0$.

In the following we will pursue the idea of constructing quantities which are covariant under the subgroup of 12d diffeomorphisms preserving the decomposition \eqref{metricdecomp}. In this way we will derive 10d eight-derivative couplings of type IIB which are automatically invariant under the $SL(2, \bR)$ symmetry\footnote{It is tempting to try and rephrase the flux-dependent type IIB action in terms of objects carrying 12d indices. One may define the four-form $\wh G_{M N R S}$ such that 
\ba \nonumber
\wh G_{m n r 1} & = H_{m n r} \, ,&  
\wh G_{m n r 2} & = F_{m n r} \, , &
\wh G_{m n r s} & = \wh G_{m n 12}  = 0 \, , & 
\ea
but these rules appear to be more ad hoc, and such a lift might require not only four-form fluxes but also a four-form potential \cite{Ferrara:1996wv, Sethi:1996es}. We will not need the knowledge of $G$-dependent, 10d $\a'^{\,3}$ terms for our purposes, and we leave their study for a future publication.}. Furthermore, it should always be kept in mind that these 10d expressions may alternatively be obtained from the F-theory limit of M-theory, even though, as opposed to the two-derivative level, ``quantum'' reductions of 11d quantities are needed \cite{Green:1997as}.

\subsection{Eight-derivative couplings of gravity and axio-dilaton}
\label{RandPsection}

 At the eight-derivative order, the 10d type IIB supersymmetric effective action for terms which depend only on the metric is known \cite{Green:1997tv,Green:1997di,Green:1997as}. It consists of a parity even-even part,  given by $f_0 t_8 t_8 R^4$, which reproduces the four-point string amplitudes, and a parity odd-odd part, given by $ f_0 \e_{10} \e_{10} R^4$, which instead gives a vanishing contribution at four-point.  However, the equivalent set of couplings depending on the derivatives of $\t$ are not fully known. Expressions for them (including the $G_3$-flux) were proposed in the past, based upon an $SL(2,\bZ)$-invariant extension of known NSNS terms \cite{Kehagias:1997cq,Kehagias:1997jg}. We will instead use the strategy outlined in section \ref{sec:2.1} to derive these couplings.

\subsubsection{\texorpdfstring{$\hat t_8 \hat t_8  \hat R^4$}{TEXT} in 12d}
\label{t8t8R4section}

The part of the 10d eight-derivative action which non-trivially contributes to the four-point amplitudes is given by $f_0 t_8 t_8 R^4$ \cite{Green:1997tv,Green:1997di,Green:1997as}. In this expression we have introduced the function $f_0$, which is the $SL(2,\bZ)$-invariant, non-holomorphic Eisenstein series of weight $3/2$ 
\be
f_0 (\t, \bar \t) \;=  \sum_{ (m,n)\neq (0,0)} \fr{\tau_2^{3/2}}{|m + n \t |^3}\, .
\ee
It is useful to note that, for large values of $\t_2$, the function $f_0$ has the expansion
\bea\label{f0expansion}
f_0 (\t, \bar \t) &=& 2 \z (3)\, \t_2^{3/2} \;+\; \fr{2\pi^2}{3}\, \tau_2^{-1/2}\; +\; \cO(e^{-\tau_2}) \, . 
\eea
We may then promote this part of the action to a quantity with 12d covariance built from the metric $ \wh g_{M N}$. This gives 
\ba\label{12dt8t8}
S^\tbthree_{t_8} = \int f_0 (\t, \bar \t)\,  \wh t_8 \wh t_8 \wh R^4  *_{10} 1\, , 
\ea
where again there is no 12d promotion of the 10d measure, and
\ba \label{eq:St8}
 \wh t_8 \wh t_8 \wh R^4 & = \wh t_{8}^{  M_1 \dots M_8} \wh t_{8 \, N_1  \dots N_8}     \wh R^{N_1 N_2}{}_{M_1 M_2} \wh R^{N_3 N_4}{}_{M_3 M_4}  \wh R^{N_5 N_6}{}_{M_5 M_6} \wh R^{N_7 N_8}{}_{M_7 M_8} \ , 
\ea
with $ \wh t_8$ as defined in appendix \ref{Conventions}.

Expanding out\footnote{Most of the computations and symbolic manipulations needed in this paper were performed in Mathematica, using the package xAct.} this expression using \eqref{metricdecomp} gives the complete $\cO(\alpha'^{\,3})$ 10d action involving gravity and axio-dilaton. The explicit expression is long and not very illuminating, but for completeness we give it in appendix \ref{BigandExplicit}.

The $t_8 t_8 R^4$ part of this expansion (see equation \eqref{St8}) is known to correspond to the four-point graviton amplitude, and so it is natural to ask if the other four-point terms in this action reproduce the remaining four-point amplitudes, which have been computed in \cite{Policastro:2006vt,Policastro:2008hg}. Many of the terms in \eqref{St8} contain more than four fields so they cannot contribute to four-point amplitudes. Removing these terms gives the simpler result
\bea
\label{St84pt}
S^\tbthree_{t_8}{}|_{ \text{4pt}} = \fr{2\zeta(3)}{g_s^{3/2}} \int \bls  t_8 t_8 R^4 + 1536 D{}^{m}P{}^{n} D{}_{m}P{}_{n} D{}^{r}\bar P{}^{s} D{}_{r}\bar P{}_{s} + 1536 D{}^{m}P{}^{n} D{}^{r}\bar P{}^{s} R{}_{m}{}^{t}{}_{r}{}^{u} R{}_{n}{}_{u}{}_{s}{}_{t} 
 \hspace{1.5cm}\nn \\ 
- 1536 D{}_{m}\bar P{}^{r} D{}^{m}P{}^{n} R{}_{n}{}^{s}{}^{t}{}^{u} R{}_{r}{}_{s}{}_{t}{}_{u} + 192 D{}_{m}\bar P{}_{n} D{}^{m}P{}^{n} R{}^{r}{}^{s}{}^{t}{}^{u} R{}_{r}{}_{s}{}_{t}{}_{u} 
+ 1536 D{}^{m}P{}^{n} D{}^{r}\bar P{}^{s} R{}_{m}{}^{t}{}_{n}{}^{u} R{}_{r}{}_{t}{}_{s}{}_{u}
\brs *_{10} 1\, .
\nn \\
\eea
In this expression we have taken the vacuum expectation value of $\tau$, because its fluctuation necessarily contributes to higher-point functions. Moreover, we have restricted to the first term of the $f_0$-expansion \eqref{f0expansion}, which corresponds to string tree-level and dominates at small values of the string coupling $g_s=e^{\langle\phi\rangle}$. This gives the constant prefactor in \eqref{St84pt}. To test whether this reproduces the correct four-point amplitudes, we now make a further expansion,  where
\ba
\t &= C_0 + i e^{-\f} \, ,& 
g_{m n} & = \h_{m n} + h_{m n} \, , 
\label{FeildExpansion}
\ea
and work to forth order in the fields $C_0$, $\f$ and $h_{m n}$. The quantities in \eqref{St84pt} are then given by
\ba
R_{m n r s} & = - \pa_{[m|} \pa_r h_{|n] s} + \pa_{[m|} \pa_s h_{|n] r} + \ldots \, , &
D_m P_n & = \fr12 ( \pa_m \pa_n \f + i \pa_m \pa_n C_0 ) + \ldots \, ,
\label{FeildStrengthExpansion}
\ea
where the dots represent terms at higher order in the fields. By going to momentum space, we can consider the individual parts of the effective action, expressed in terms of Mandelstam variables. Ignoring the constant prefactor in \eqref{St84pt}, this gives
\ba
\cL^\tbthree_{t_8}{}^{ \f^2 R^2 } &= 12 t^2 u^2 \til \f(k_1) \til \f(k_2) \til h_{m n} ( k_3 ) \til h^{m n} ( k_4 ) + \ldots +  \text{perms} \, , 
\label{t8f2R24pt}
\ea
\ba
\cL^\tbthree_{t_8}{}^{ C_0^2 R^2 } &= 12 t^2 u^2 \til C_0(k_1) \til C_0(k_2) \til h_{m n} ( k_3 ) \til h^{m n} ( k_4 ) + \ldots + \text{perms}\, , 
\label{t8C2R24pt}
\ea
\ba
\cL^\tbthree_{t_8}{}^{ \f^4 } &= 2 ( s^4 + t^4 + u^4 ) \til \f(k_1) \til \f(k_2)  \til \f ( k_3 ) \til \f ( k_4 ) + \text{perms} \, ,
\label{t8f44pt}
\ea
\ba
\cL^\tbthree_{t_8}{}^{ \f^2 C_0^2 } &= 12 ( -s^4 + t^4 + u^4 ) \til \f(k_1) \til \f(k_2) \til C_0 ( k_3 ) \til C_0( k_4 ) + \text{perms} \, , 
\label{t8f2C24pt}
\ea
\ba
\cL^\tbthree_{t_8}{}^{ C_0^4 } &= 2  ( s^4 + t^4 + u^4 )  \til C_0(k_1) \til C_0(k_2) \til C_0 ( k_3 ) \til C_0 ( k_4 ) + \text{perms} \, , 
\label{t8C44pt}
\ea
where, as in  \cite{Policastro:2008hg}, the ellipses hide terms with contractions between $k_m$ and $h_{m n}$, and the extra permutations give the terms with all possible exchanges of $k_1$, $k_2$, $k_3$ and $k_4$. These expressions represent the various parts of the momentum-space-transformed integrand in \eqref{St84pt}. 
Comparing them with the analogous expressions for the four-point string amplitudes given in \cite{Policastro:2008hg}, we see that \eqref{St8}, and thus the torus reduction of \eqref{12dt8t8}, exactly reproduces the known four-point results\footnote{To see this explicitly, it is convenient to use the identity
\ba
 ( s^4 + t^4 + u^4 ) \til \f(k_1) \til \f(k_2)  \til \f ( k_3 ) \til \f ( k_4 ) + \text{perms} =  3 ( - s^4 + t^4 + u^4 ) \til \f(k_1) \til \f(k_2)  \til \f ( k_3 ) \til \f ( k_4 ) + \text{perms} \, , 
\ea
where the difference is re-absorbed by the set of extra permutations. Similar expression holds for terms with four $\til C_0$'s. 
}. 

A similar analysis also confirms the observation of \cite{Policastro:2008hg} that, defining 
\ba
\d DP_{m n}{}^{rs} = \d_{[m}{}^{[r} D_{n]} P^{s]}    \, , 
\ea
\eqref{t8f2R24pt} and \eqref{t8C2R24pt} can be compactly written as $t_8 t_8 R^2 \d DP \d D \bar P$, but that \eqref{t8f2C24pt} is not compatible with the structure $t_8 t_8 \d DP^2 \d D \bar P^2$.

Strictly speaking, the above represents a sufficient comparison only for \eqref{t8f44pt} to \eqref{t8C44pt}. But to check that our result agrees with \cite{Policastro:2008hg} for \eqref{t8f2R24pt} and \eqref{t8C2R24pt}, we should also match the dotted part, which is not given explicitly in  \cite{Policastro:2008hg}. 
Fortunately, in these terms the number of explicit derivatives is lower. Hence we can directly match the $R^2 DP D\bar P$ terms of \eqref{St84pt} with those coming from $t_8 t_8 R^2 \d DP \d D \bar P$, which was proven in \cite{Policastro:2008hg} to correctly package this sector of the four-point effective action (two gravitons + two axio-dilatons). This match can be demonstrated to hold if we make use of a large number of total-derivative identities and of the field equations at lowest order in $\alpr$ and in points. The latter imply that $ D_m P^m = R_{m n} = 0$, as is proven in detail in appendix \ref{RedefinitionsandHigherDerivativeActions}. Altogether, we are able to show that
\ba
24  \int t_8 t_8 R^2 \d DP \d D \bar P *_{10} 1 &  = \int \bls  1536 D{}^{m}P{}^{n} D{}^{r}\bar P{}^{s} R{}_{m}{}^{t}{}_{r}{}^{u} R{}_{n}{}_{u}{}_{s}{}_{t} 
- 1536 D{}_{m}\bar P{}^{r} D{}^{m}P{}^{n} R{}_{n}{}^{s}{}^{t}{}^{u} R{}_{r}{}_{s}{}_{t}{}_{u} 
\nn \\ & 
+ 192 D{}_{m}\bar P{}_{n} D{}^{m}P{}^{n} R{}^{r}{}^{s}{}^{t}{}^{u} R{}_{r}{}_{s}{}_{t}{}_{u} 
+ 1536 D{}^{m}P{}^{n} D{}^{r}\bar P{}^{s} R{}_{m}{}^{t}{}_{n}{}^{u} R{}_{r}{}_{t}{}_{s}{}_{u}
\brs *_{10} 1
\nn \\ & 
 + \cO ( R_{m n}) + \cO ( D_m P^m ) \, . 
\ea
This fully confirms that our result \eqref{St84pt} correctly reproduces the known four-point effective action.

\subsubsection{\texorpdfstring{$ {\hat \e}_{12} { \hat \e}_{12} { \hat R^4}$}{TEXT} in 12d}
\label{e12e12R4section}

Next, let us consider the equivalent generalization of the $ f_0 \e_{10} \e_{10} R^4$ terms present in 10d. Promoting this part of the 10d action to a quantity with 12d covariance, we find
\ba\label{12de12e12}
S^\tbthree_{\e_{12}} &= \fr1{96} \int  f_0 (\t, \bar \t) \, \wh \e_{12} \wh \e_{12} \wh R^4  *_{10} 1 \, , 
\ea
where
\ba\label{eq:epepR4}
 \wh \e_{12} \wh \e_{12} \wh R^4 &=  \epsilon^{R_1 R_2 R_3 R_4 M_1\ldots M_{8} }  \epsilon_{R_1 R_2 R_3 R_4  N_1 \ldots N_{8}} \wh R^{N_1 N_2}{}_{M_1 M_2} \wh R^{N_3 N_4}{}_{M_3 M_4}  \wh R^{N_5 N_6}{}_{M_5 M_6} \wh R^{N_7 N_8}{}_{M_7 M_8} \, . 
\ea
Expanding out this expression gives a formula which is too unwieldy to be placed in the main text. Yet, for completeness, we provide the full expansion in appendix  \eqref{BigandExplicit}.

As before, we may perform our four-point analysis by restricting to only the small part of the action \eqref{eq:epepR4-expand} which can contribute. This is given by 
\ba
S^\tbthree_{\e_{12}}|_{ \text{4pt}} &=  \fr{\zeta(3)}{ 4 g_s^{3/2}} \int \bls  \e_{10} \e_{10}  R^4
-192 D{}_{m}\bar P{}_{n} D{}^{m}P{}^{n} R{}^{2}
 + 192 D{}^{m}P{}_{m} D{}^{n}\bar P{}_{n} R{}^{2}
  - 768 D{}^{m}P{}^{n} D{}^{r}\bar P{}_{r} R R{}_{m}{}_{n} 
     \nn \\& 
+ 768 D{}^{m}P{}^{n} D{}^{r}\bar P{}^{s} R R{}_{m}{}_{r}{}_{n}{}_{s} 
+ 1536 D{}_{m}\bar P{}^{r} D{}^{m}P{}^{n} R R{}_{n}{}_{r} 
- 768 D{}^{m}P{}_{m} D{}^{n}\bar P{}^{r} R R{}_{n}{}_{r}
   \nn \\& 
 + 3072 D{}^{m}P{}^{n} D{}^{r}\bar P{}^{s} R{}_{m}{}^{t} R{}_{n}{}_{r}{}_{s}{}_{t}
  - 1536 D{}^{m}P{}^{n} D{}^{r}\bar P{}^{s} R{}_{m}{}_{r} R{}_{n}{}_{s} 
  + 1536 D{}^{m}P{}^{n} D{}^{r}\bar P{}_{r} R{}_{m}{}^{s} R{}_{n}{}_{s} 
     \nn \\& 
  + 768 D{}^{m}P{}^{n} D{}^{r}\bar P{}^{s} R{}_{m}{}_{r}{}^{t}{}^{u} R{}_{n}{}_{s}{}_{t}{}_{u}
   - 768 D{}^{m}P{}^{n} D{}^{r}\bar P{}_{r} R{}_{m}{}^{s}{}^{t}{}^{u} R{}_{n}{}_{s}{}_{t}{}_{u} 
   + 1536 D{}^{m}P{}^{n} D{}^{r}\bar P{}^{s} R{}_{m}{}^{t}{}_{r}{}^{u} R{}_{n}{}_{t}{}_{s}{}_{u} 
      \nn \\& 
   + 1536 D{}^{m}P{}^{n} D{}^{r}\bar P{}^{s} R{}_{m}{}_{n} R{}_{r}{}_{s} 
   - 3072 D{}_{m}\bar P{}^{r} D{}^{m}P{}^{n} R{}_{n}{}^{s} R{}_{r}{}_{s} 
   + 1536 D{}^{m}P{}_{m} D{}^{n}\bar P{}^{r} R{}_{n}{}^{s} R{}_{r}{}_{s} 
      \nn \\& 
   + 768 D{}_{m}\bar P{}_{n} D{}^{m}P{}^{n} R{}^{r}{}^{s} R{}_{r}{}_{s} 
   - 768 D{}^{m}P{}_{m} D{}^{n}\bar P{}_{n} R{}^{r}{}^{s} R{}_{r}{}_{s} 
   + 1536 D{}_{m}\bar P{}^{r} D{}^{m}P{}^{n} R{}_{n}{}^{s}{}^{t}{}^{u} R{}_{r}{}_{s}{}_{t}{}_{u} 
      \nn \\& 
   - 768 D{}^{m}P{}_{m} D{}^{n}\bar P{}^{r} R{}_{n}{}^{s}{}^{t}{}^{u} R{}_{r}{}_{s}{}_{t}{}_{u} 
   - 192 D{}_{m}\bar P{}_{n} D{}^{m}P{}^{n} R{}^{r}{}^{s}{}^{t}{}^{u} R{}_{r}{}_{s}{}_{t}{}_{u} 
   + 192 D{}^{m}P{}_{m} D{}^{n}\bar P{}_{n} R{}^{r}{}^{s}{}^{t}{}^{u} R{}_{r}{}_{s}{}_{t}{}_{u} 
      \nn \\& 
   - 3072 D{}^{m}P{}^{n} D{}^{r}\bar P{}^{s} R{}_{m}{}_{s}{}_{n}{}_{t} R{}_{r}{}^{t} 
   - 1536 D{}^{m}P{}^{n} D{}^{r}\bar P{}^{s} R{}_{m}{}^{t}{}_{n}{}^{u} R{}_{r}{}_{t}{}_{s}{}_{u} 
   + 1536 D{}^{m}P{}^{n} D{}^{r}\bar P{}_{r} R{}_{m}{}_{s}{}_{n}{}_{t} R{}^{s}{}^{t} 
      \nn \\& 
   - 3072 D{}_{m}\bar P{}^{r} D{}^{m}P{}^{n} R{}_{n}{}_{s}{}_{r}{}_{t} R{}^{s}{}^{t} 
   + 1536 D{}^{m}P{}_{m} D{}^{n}\bar P{}^{r} R{}_{n}{}_{s}{}_{r}{}_{t} R{}^{s}{}^{t} \brs *_{10} 1\, .
\ea 
This expression is not exactly short and elegant, but it can be subjected to a basic sanity test. Indeed, we may expand the fields as described in \eqref{FeildExpansion} and \eqref{FeildStrengthExpansion} and go to momentum space, where we find that 
\ba
\cL^\tbthree_{\e_{12}}{}^{ \f^2 R^2 } = \cL^\tbthree_{\e_{12}}{}^{ C_0^2 R^2 } = \cL^\tbthree_{\e_{12}}{}^{ \f^4 } = 
\cL^\tbthree_{\e_{12}}{}^{ \f^2 C_0^2 } =  
\cL^\tbthree_{\e_{12}}{}^{ C_0^4 } = 0 \, . 
\ea
This is consistent with the results of \cite{Policastro:2008hg}, and shows that, like the standard $\e_{10} \e_{10} R^4$ part, also its axio-dilaton completion leads to a vanishing four-point amplitude.

\section{Four-dimensional $\cN=1$ compactifications}
\label{sec:4d}

In this section we would like to take the 12d framework one step further. We want to compactify the purely gravitational sector of the 12d action, corrected at order $\alpha'^{\,3}$ as discussed in the previous section, down to 4d on an elliptically fibered CY fourfold, and study the consequences on the ensuing $\cN=1$ effective action at two-derivative level. Anticipating the result, this is going to give us a new, genuinely $\cN=1$ correction to the 4d effective action: This correction is due to backreaction effects of 7-branes on the closed string background, and therefore modify the long-known correction \cite{Antoniadis:1997eg,Becker:2002nn}, which instead originates from the $\cN=2$ configurations of closed strings living on Ricci-flat spaces.

First, let us consider the two-derivative action \eqref{R12act}
\ba
S^\tbzero = \fr1{ (2 \pi)^7 \alpha'^{\,4}} \int  \wh R *_{10} 1 \, ,
\label{R12actWithCoef}
\ea
and reduce it to 4d using the following ansatz for the 12d metric:
\ba
d \wh s ^2 = g_{\m \n} dx^\m dx^\n + 2 g_{a \bar b} dy^a dy^{\bar b}\,.
\label{12d4dMetricDecomp}
\ea
In this expression $g_{\m \n}$ is the metric on the 4d external space $\mathbb{R}^{1,3}$ and $g_{a \bar b}$ is the metric on the CY fourfold $X_4$, which we take to be a smooth elliptic fibration over a base $B_3$, a six-dimensional K\"ahler manifold. This represents an expansion about a solution to the lowest order 10d field equations, where the external space metric is that of flat Minkowski space. The $*_{10}1 $ in \eqref{R12actWithCoef} represents the volume form on the 10d space given by the product of the 4d external space and the base\footnote{More precisely, here we mean the zero-section of the fibration, which is a holomorphic 6-cycle, diffeomorphic to the base. In this paper we only consider fibrations admitting such a zero-section.} $B_3$. After Kaluza-Klein reduction, we simply get
\begin{equation}
S^{(0)}_{\rm(4d)}=\frac{1}{2\pi \alpha'}\int \mathcal{V}_b \, R_{\rm(4d)}\,*_4 1\,,
\label{4dLowestAct}
\end{equation}
where $R_{\rm(4d)}$ is the Ricci scalar in 4d and $\mathcal{V}_b$ is the classical volume of $B_3$ in units of $2\pi\sqrt{\alpha'}$. Had we started with $*_{12}1$ in 12d, we would have seen more terms in \eqref{4dLowestAct} due to the non-triviality of the fibration. These additional terms are precisely the ones which in the F/M-theory duality are killed by the F-theory limit \cite{Grimm:2010ks}. In the present 12d framework, instead, they are automatically excluded by the use of $*_{10}1$ in \eqref{R12actWithCoef} and the restriction to the base which we have performed. 

In order to put the action \eqref{4dLowestAct} into the Einstein frame, a Weyl rescaling must be performed where $g_{\m \n} \ra g_{\m \n}/\cV_b$. In what follows we will study how the higher-derivative terms modify such rescaling. We emphasize that corrections to the Weyl rescaling translate into corrections to the K\"ahler potential, but there might be yet further corrections to the latter, which could be equally relevant for phenomenology \cite{Berg:2014ama}. We would also like to stress that the process of deriving a correction to the K\"ahler potential from a correction to the Weyl rescaling factor is well-established in situations where there is a supergravity description for the higher-dimensional theory\footnote{We thank the referee for raising this point.}. In this regard, we are adopting the logic of the M-theory approach to F-theory,  according to which the higher-dimensional theory in question is the 11d supergravity. Thus, all $g_s$ effects  are codified by the curvature of the compactification manifold (the elliptic fibration), and the source terms are conveniently ``geometrized''. Where our method slightly differs from the M-theory approach is in that it does not require the zero-size limit of the fiber. This concerns only how one treats the K\"ahler structure of the internal manifold. As far as its complex structure is concerned (which is what really controls $g_s$ effects), our approach is totally equivalent to the traditional M-theory one.

Let us now move on to the action at order $\alpha'^{\,3}$. Combining \eqref{12dt8t8} and \eqref{12de12e12}, we have
\be\label{CompleteS3}
S^{(3)}=\frac{1}{(2\pi)^7\cdot3\cdot2^{11} \cdot \alpha'} \int f_0(\tau,\bar\tau)\, (t_8t_8+\tfrac{1}{96}\epsilon_{12}\epsilon_{12}) \hat{R}^4\,*_{10}1\,.
\ee
As with the two-derivative part, the action in \eqref{CompleteS3} has to be thought of as the integral over a 10d slice of the 12d space. To avoid cluttering the notation, we omit the symbol indicating the restriction to this slice.
By decomposing the part of the 12d integrand which contributes to the 4d Weyl rescaling we find
\bea
\int f_0 \fr{1}{96} \wh \e_{12} \wh \e_{12} \wh R^4 *_{10} 1 &=& - 768 (2\pi)^3 \int R_{\rm(4d)} *_4 1 \int_{B_3} f_0 *_8 ( J \we c_3(X_4) ) *_6 1
 + \ldots\,,
\label{WeylCorrection1}
\eea
where $*_8$ represents the Hodge dual on the CY fourfold, $*_6 1$ is the volume form of $B_3$ and the ellipses indicate terms which are not linear in the 4d Ricci scalar. In this expression we have also introduced $c_3(X_4)$ which is the third Chern form on the CY fourfold given by \eqref{c3def}. It is important to note that this is a particular representative of the third Chern class of the tangent bundle of $X_4$ and does not represent the class as a whole, because exact shifts in $c_3(X_4)$ modify the result \eqref{WeylCorrection1}. If we then demand, as in section \ref{sec:10d}, that all derivatives in fiber directions vanish and use the K\"ahler property of the CY metric, we see that $c_3(X_4)$ only has legs in base directions and so is a top-form on $B_3$, given by the restriction $c_3(X_4)|_{B_3}$. This means that we may rewrite the r.h.s. of \eqref{WeylCorrection1} as 
\ba
- 768 (2\pi)^3 \int R_{\rm(4d)} *_4 1 \int_{B_3} f_0 *_8 ( J \we  *_6 1) c_3(X_4) |_{B_3} \,.
\label{WeylCorrection2}
\ea
Next, we make a decomposition of the CY fourfold K\"ahler form $J$ into $J = J_f + J_b$, where  $J_b$ is the K\"ahler form on the base and $J_f$ the fixed volume of the fiber. As the space is a non-trivial fibration, $J_f \we J_f$ does not vanish, but instead is cohomologically equivalent to a quantity proportional to $c_1(B_3) \we J_f$, where $c_1(B_3)$ is the first Chern class of $B_3$. However, for dimensional reasons, these non-linear powers of $J_f$ comes with extra factors of $\alpha'$, so that $ *_8 ( J \we  *_6 1)  = 1 + \cO(\alpha') $.
All of the $\alpha'$ corrections in this expansion give contributions to our result which are higher-order than $\alpha'^{\,3}$, and thus may be neglected in our analysis. Therefore, combining \eqref{WeylCorrection2} with the classical action \eqref{4dLowestAct}, we find the following corrected factor of the 4d Weyl rescaling:
\bea
S_{\rm(4d)}&=&\frac{1}{2\pi \alpha'}\int\,\left(\mathcal{V}_b-\frac{1}{64\pi^3}\int_{B_3} f_0(\tau,\bar\tau) \,c_3(X_4) |_{B_3} \right)\, R_{\rm(4d)}\,*_4 1\,,
\label{WeylCorrection3}
\eea

For backgrounds with constant axio-dilaton (trivial elliptic fibrations, no 7-branes), the base is a CY threefold and formula \eqref{WeylCorrection3} exactly reproduces the known $\cN=2$ correction of \cite{Antoniadis:1997eg,Becker:2002nn}. However, if the vacuum expectation value of $\tau$ depends on the coordinates of the base, which is the case in $\cN=1$ backgrounds, the correction in \eqref{WeylCorrection3} is non-topological, and the integral is difficult to perform. 

One might worry that this correction diverges, but we argue that this does not happen. Indeed, $f_0$ is a real analytic function on the upper half $\tau$ plane $\mathbb{H}$. Moreover, since it is $SL(2,\mathbb{Z})$-invariant, its domain of definition can be taken to be the fundamental region $\mathbb{H}/SL(2,\mathbb{Z})$, and its only pole is at imaginary infinity, which is the value of $\tau$ at D7-brane locations. Hence we need to check that, despite this divergence, the integral in \eqref{WeylCorrection3} is still finite. On the basis of arguments analogous to those of \cite{Greene:1989ya}, we argue that the Riemann tensors contained in $c_3(X_4)$ do not cause problems, so we concentrate just on $f_0$. Consider then a disk of unit radius around a D7-brane. The profile of $\tau$ in this disk is given by
\be\label{LocalTau}
\tau=\tau^{(0)} + \frac{1}{2\pi i}\log(z)\,,
\ee
where $z$ is the complex coordinate on the disk and $\tau^{(0)}$ is the ``asymptotic'' axio-dilaton, i.e. $(\tau^{(0)}_2)^{-1}=g_s$ is the stabilized value of the string coupling. Using the fact that $\tau$ diverges as $-i\log(|z|)$ when approaching the D7 at $z=0$, and that, for large $\tau_2$, $f_0$ goes like $(\tau_2)^{3/2}$ (see equation \eqref{f0expansion}), it is easy to verify that the integral of $f_0$ on the disk is indeed finite.

At this point, we would like to make a brief consideration on the validity of our approach, based, as remarked above, on having a supergravity description in 11d. The main possible objection could be that there are regions of high curvature, as a result of the presence of regions of strong coupling from the type IIB perspective, which may invalidate the 11d supergravity approximation. The key feature, which we believe helps us avoiding this issue, is the $SL(2,\mathbb{Z})$ invariance of the higher-derivative couplings we are discussing. Even though in F-theory a global $SL(2,\mathbb{Z})$ rotation to weak string coupling cannot be performed, we can still work patch-wise on the string internal manifold $B_3$, and in each patch use a frame in which the string coupling is small and the supergravity analysis can be trusted. We can do that precisely because the integrand which expresses our correction in formula \eqref{WeylCorrection3} is the same in every frame (being $SL(2,\mathbb{Z})$-invariant). We believe that the fact that the integral in \eqref{WeylCorrection3} is not divergent is a strong indication that such a ``patch-wise integration'' is sensible. To conclude, we notice that the problem already arises at lowest order in $\alpha'$ and  that the same argument may be used in showing that everything should work out correctly.  In that case, the statement is that at strong coupling the K\"ahler potential should be proportional to the logarithm of the classical volume of the base of the elliptic fibration (which replaces the classical volume of the Calabi-Yau threefold double cover at weak coupling) \cite{Grimm:2010ks}. Indeed, the classical volume of the base (in the Einstein frame) is also an $SL(2,\mathbb{Z})$-invariant quantity.

Formula \eqref{WeylCorrection3} can be drastically simplified by going to the Sen weak coupling limit of F-theory \cite{Sen:1996vd}. Sen's prescription consists in moving to a region of the complex structure moduli space of the CY fourfold, where none of the monodromies acting on $\tau$ involves the string coupling $(\tau_2)^{-1}$, which thus can be kept small in a globally well-defined way. One can describe this weakly coupled physics by means of type IIB string theory compactified on an orientifolded CY threefold $X_3$. The latter has the property of being the double cover of $B_3$, branched along the divisor wrapped by an O7-plane. The Poincar\'e dual of this divisor is the pull-back class $D_{\rm O7}\equiv\pi^*c_1(B_3)$, under the projection map $\pi:X_3\to B_3$. Moreover, there is a single, orientifold invariant D7-brane, wrapping the divisor of class $8\,\pi^*c_1(B_3)$.

In this regime, where we take $g_s\ll1$, we may consistently neglect the varying part of $\tau_2$ in equation \eqref{LocalTau}, and therefore pull $f_0$ out of the integral. We must be careful, though, because at $\cO(e^{-1/g_s})$ distances from the D7-brane, the varying contribution in \eqref{LocalTau} becomes comparable to the constant part. However, it is easy to check by explicit computation, that the dominant term of the integral goes like $g_s^{-3/2}$ and is just given by the constant part of \eqref{LocalTau}. Hence, for very small string coupling, we have
\be\label{WeylCorrection4}
S_{\rm(4d)}=\frac{1}{2\pi \alpha'}\int\,\left(\mathcal{V}_b-\frac{\zeta(3)}{32\pi^3\,g_s^{3/2}}\int_{B_3} c_3(X_4) |_{B_3} \right)\, R_{\rm(4d)}\,*_4 1 \quad+\quad \cO(g_s^{-1/2})\,.
\ee
At this point our correction becomes topological, and the quantity $c_3(X_4)|_{B_3}$ can as well be regarded as a cohomology class. Notice that the next-to-leading order correction appears already at $\cO(g_s^{-1/2})$, and not at $\cO(g_s^{1/2})$, as was the case for \cite{Antoniadis:1997eg,Becker:2002nn}. This was expected, due to the presence of the D7-brane and the O7-plane, which allow for non-trivial contributions of both open string and non-orientable string diagrams.

We can simplify \eqref{WeylCorrection4} further, by expressing $c_3(X_4)|_{B_3}$ in terms of Chern classes of the CY threefold $X_3$. Using adjunction formulae, it is easy to show that, in cohomology,
\ba
c_3(X_4)|_{B_3} &=c_3(B_3)-c_1(B_3)c_2(B_3)\, ,   & 
\pi^*(c_3(B_3)-c_1(B_3)c_2(B_3)-2 c_1^3(B_3)) & = c_3(X_3)\,.
\ea
Substituting this into \eqref{WeylCorrection4} and turning integrals over $B_3$ into integrals over $X_3$, we find the following correction to the classical volume $\cV_3$ of the CY threefold:
\bea\label{quantumVolume3}
\tilde{\mathcal{V}}_3 & = & \mathcal{V}_3 \;-\; \frac{\zeta(3)}{32\pi^3\,g_s^{3/2}} \,\left(\chi(X_3)+2\int_{X_3}D_{\rm O7}^3\right)\quad+\quad \cO(g_s^{-1/2})\,,
\eea
where we recall that $D_{\rm O7}\equiv\pi^*c_1(B_3)$ is the class Poincar\'e dual to the O7-plane in $X_3$.
The first correction term in \eqref{quantumVolume3} is the one of
\cite{Antoniadis:1997eg,Becker:2002nn}. The second term is new, and it is a genuinely $\cN=1$ correction. From a string point of view, it should arise from tree-level closed string scattering in this CY orientifold background.

We might be concerned that the correction to the Weyl rescaling \eqref{quantumVolume3} does not properly take into account the contribution from source terms which should be present in the type IIB effective action. Performing the reduction of our proposed 12d action, using the metric decomposition \eqref{metricdecomp}, we get the bulk IIB action. But, when the profile for $\t$ becomes non-trivial, the latter is certainly corrected by source terms. However, it is important to note that \eqref{metricdecomp} is only well-defined away from 7-brane sources, around which $\t$ undergoes non-trivial monodromies. In this section, instead, our reduction is performed with the metric decomposition \eqref{12d4dMetricDecomp}, which should be true globally. If the CY metric in \eqref{12d4dMetricDecomp} is smooth, we believe that the difference from \eqref{metricdecomp} should be such that the source terms are already taken into account by our 12d expressions, and the full action is \eqref{R12actWithCoef} plus \eqref{CompleteS3} to order $\alpha'^{\,3}$. 
If the F-theory CY fourfold is singular, which is always the case when we have non-trivial gauge groups at low energy, it may seem impossible to perform our analysis, as in our 12d approach we are not allowed to perform resolutions. However, if gauge groups are engineered via the standard Tate algorithm \cite{Bershadsky:1996nh}, we believe that our correction should persist even if it gets modified by further additional terms. Indeed, singularities would arise away\footnote{Embedding the fiber in W$\mathbb{P}^2_{231}$ with homogeneous coordinates $X,Y,Z$, the zero section is at the point $Z=0$, whereas singularities are usually enforced at $X=Y=0$.} from the zero-section, where our correction \eqref{WeylCorrection3} lives, and so would be unable to alter it. In cases where extra source terms should be added to our proposed 12d action, these may result in additional terms to those shown in \eqref{WeylCorrection3}, and consequently to those in \eqref{quantumVolume3}.

Let us end by noting a few caveats related to this correction. Firstly, we assumed that the full set of eight-derivative, 10d terms containing only gravity and axio-dilaton is given by \eqref{CompleteS3} at all orders in the fields. In section \ref{sec:10d} we have demonstrated that this is true at four-point, but whether \eqref{CompleteS3} is enough to correctly account also for higher-point amplitudes remains to be seen. Secondly, the effect of corrections to the vacuum solution that will be induced by the higher-derivative terms we have included has not been taken into account here. These terms will modify the 12d equations of motion and can make the solution deviate from the CY one, which is present at lowest order. These sorts of corrections include those associated with warping and changes to the internal space metric that can significantly alter the ansatz \eqref{12d4dMetricDecomp} at order $\a^{\pr \, 3}$, as seen in \cite{Nemeschansky:1986yx,Freeman:1986br,Freeman:1986zh,Grimm:2014xva,Grimm:2014efa}. These effects could also give contributions which are equally relevant to those we compute here in the Weyl rescaling, and may modify our result. Finally, while we have computed a contribution to the Weyl rescaling, this does not fully determine the correction to the K\"ahler potential. One would need to also compute corrections to the kinetic terms for the moduli appearing in the dimensional reduction \cite{Berg:2014ama}, in order to draw precise conclusions about the corrected K\"ahler potential.

\vspace{-.0029cm}

\section*{Acknowledgments}

We would like to thank T. Grimm for initial collaboration and many useful conversations. We also express our gratitude to A. Collinucci, I. Garc\'ia-Etxebarria, L. Martucci, D. Morrison,  G. Policastro,  S. Sasmal, D. Tsimpis, P. Vanhove and especially to M. Gra\~na for helpful discussions and valuable insights. This work was supported in part by the Agence Nationale de la Recherche under the grant 12-BS05-003-01 (RM),  by a grant of the Max Planck Society (TP) and by the ERC Starting Grant 259133 - ObservableString (RS).

\begin{appendix}
\vspace{2cm} 

\section{Conventions }
\label{Conventions}

We adopt the following conventions for the Christoffel symbols and the Riemann tensor 
\ba
\wh \G^R{}_{M N} & = \fr12 \wh g^{RS} ( \pa_{M} \wh g_{N S} + \pa_N \wh g_{M S} - \pa_S \wh g_{M N}  ) \, , &
\wh R_{M N} & = \wh R^R{}_{M R N} \, , \nn \\
\wh R^{M}{}_{N R S} &= \pa_R \wh \G^M{}_{S N}  - \pa_{S} \wh \G^M{}_{R N} + \wh \G^M{}_{R  T} \wh \G^T{}_{S N} - \wh \G^M{}_{ST}  \wh \G^T{}_{R N} \,, &
\wh R & = \wh R_{M N}  \wh g^{M N} \, . 
\ea
The tensor $\wh t_8 $ is given by the product of metrics 
\ba \label{def-t8}
\hat t_8^{N_1\dots N_8}   &= \fr{1}{16} \big( -  2 \left(   \wh g^{ N_1 N_3  }\wh g^{  N_2  N_4  }\wh g^{ N_5   N_7  }\wh g^{ N_6 N_8  } 
 + \wh g^{ N_1 N_5  }\wh g^{ N_2 N_6  }\wh g^{ N_3   N_7  }\wh g^{  N_4   N_8   }
 +  \wh g^{ N_1 N_7  }\wh g^{ N_2 N_8  }\wh g^{ N_3   N_5  }\wh g^{  N_4 N_6   }  \right) \nn \\
 & \quad +
 8 \left(  \wh g^{  N_2     N_3   }\wh g^{ N_4    N_5  }\wh g^{ N_6    N_7  }\wh g^{ N_8   N_1   } 
  +\wh g^{  N_2     N_5   }\wh g^{ N_6    N_3  }\wh g^{ N_4    N_7  }\wh g^{ N_8   N_1   } 
  +   \wh g^{  N_2     N_5   }\wh g^{ N_6    N_7  }\wh g^{ N_8    N_3  }\wh g^{ N_4  N_1   } 
\right) \nn \\
& \quad - (N_1 \lra N_2) -( N_3 \lra N_4) - (N_5 \lra N_6) - (N_7 \lra N_8) \big) \,. 
\ea
The epsilon tensor is defined such that
\ba
\wh \epsilon^{R_1\cdots R_p N_{1 }\ldots N_{d-p}}\epsilon_{R_1 \ldots R_p M_{1} \ldots M_{d-p}} &= (-1)^s (d-p)! p! 
\delta^{N_{1}}{}_{[M_{1}} \ldots \delta^{N_{d-p}}{}_{M_{d-p}]} \,, 
\ea
where $s=0$ for a Riemannian signature metric and $s=1$ for a Lorentzian signature metric. 
Equivalent definitions are used for the relevant 10d quantities.

In the reduction of section \ref{sec:4d} we have introduced $c_3(X_4)$, which is the third Chern form associated to the CY metric $g_{a \bar b}$. This is given by 
\ba
c_3 (X_4) & = -\fr{i}{3 (2\pi)^3 } R_{a}{}^b \we R_{b}{}^{c} \we R_{c}{}^a \, .
\label{c3def}
\ea


\section{Higher-derivative actions}
\label{HigherDerivativeActions}
In this appendix we show some redefinitions useful in writing higher-derivative couplings in the effective actions, as well as a collection of formulae that are too big to fit in the main text.
\subsection{Redefinitions }
\label{RedefinitionsandHigherDerivativeActions}

Let us consider an action known at second order in derivatives such that
\ba
S = S^\tbzero ( g_{m n}, \t) \, . 
\ea
When the fields undergo a small shift, the induced shift of the action is given by 
\ba
\d S = \int  ( E^\tbzero (g)^{m n} \d g_{m n}   + E^\tbzero (\t) \d \t + E^\tbzero (\bar \t) \d \bar \t ) *_{10} 1 \, , 
\ea
where
\ba
 E^\tbzero (g)^{m n} &= \fr{1}{\sqrt {-g}} \fr{ \d S^\tbzero}{\d g_{m n}}  \, , & 
 E^\tbzero (\t) &= \fr{1}{\sqrt {-g}}  \fr{ \d S^\tbzero}{\d \t } \, , 
\ea
represent the quantities which vanish on the equations of motion. 

If we now include a set of higher-derivative terms in the action suppressed by $\alpr$, the total action becomes
\ba
S = S^\tbzero ( g_{m n}, \t)  + \alpr^{\,3} S^\tbthree ( g_{m n}, \t) \, . 
\ea
Next, let us consider a redefinition of the fields such that 
\ba
\t &\ra \t + \alpr^{\,3} T \,, &
g_{mn} &\ra g_{mn} + \alpr^{\,3} G_{m n} \,, 
\ea
The action written in terms of the new fields is then given by
\ba
S = S^\tbzero ( g_{m n}, \t)  + \alpr^{\,3} S^\tbthree ( g_{m n}, \t)  +   \alpr^{\,3}  \int (  E^\tbzero (g)^{m n} G_{m n}   + E^\tbzero (\t) T +   E^\tbzero (\bar \t) \bar T ) *_{10}1  \,.
\ea
This shifts the higher-derivative part without shifting the lowest-order part. Thus we see that the higher-derivative part of the action is defined only up to terms which vanish on the lowest-order field equations, as such terms may always be absorbed by an $\alpr$-dependent field redefinition. 

In our case 
\ba
S^\tbzero = \int  ( R - 2 P_m \bar P^m ) *_{10}1 \, , 
\ea
so that 
\ba
E^\tbzero (g)_{mn} &= R_{m n} - \fr12 R g_{m n}  - 2  P_{(m} \bar P_{n)} + g_{m n} P_r \bar P^r \,,  & 
E^\tbzero (\t) &= -\fr{2 i}{\t^2} D_m P^m  \,, 
\ea
which imply that 
\ba
D_m P^m &= 0 \, ,& 
R_{m n} &= 2 P_{(m} \bar P_{n)} \,. 
\ea

In the four-point effective action we consider, substituting in $R_{m n} = 2 P_{(m} \bar P_{n)} $ will automatically produce a quantity which is  five-point or higher, as $R$ and $P$ are both linear in fields at lowest order. We may therefore treat the second equation as $R_{m n} = 0$ when considering the four-point effective action. 

\subsection{Intermediate results}
\label{BigandExplicit}

Here we collect two formulae that may turn useful, but due to their size are unpleasant to include in the main body of the paper.

The full expansion of \eqref{12dt8t8}, using the 12d metric decomposition \eqref{metricdecomp}, yields
\ba
S^\tbthree_{t_8} &= \int f_0(\tau,\bar{\tau})\, \bls  t_8 t_8 R^4 
+ 1536 D{}^{m}P{}^{n} D{}_{m}P{}_{n} D{}^{r}\bar P{}^{s} D{}_{r}\bar P{}_{s} 
- 3072 \bar P{}^{n} \bar P{}^{r} D{}_{n}P{}^{s} D{}_{r}\bar P{}_{s} P{}^{m} P{}_{m} 
\nn \\ & 
+ 1536 \bar P{}^{n} \bar P{}_{n} D{}_{r}\bar P{}_{s} D{}^{r}P{}^{s} P{}^{m} P{}_{m} 
- 3072 \bar P{}^{r} \bar P{}^{s} D{}_{m}P{}_{r} D{}_{n}\bar P{}_{s} P{}^{m} P{}^{n} 
- 3072 \bar P{}^{r} \bar P{}_{r} D{}_{m}P{}^{s} D{}_{n}\bar P{}_{s} P{}^{m} P{}^{n} 
\nn \\ &
+ 7680 \bar P{}^{r} \bar P{}^{s} D{}_{m}P{}_{n} D{}_{r}\bar P{}_{s} P{}^{m} P{}^{n} 
+ 6144 \bar P{}_{m} \bar P{}^{r} D{}_{n}P{}^{s} D{}_{r}\bar P{}_{s} P{}^{m} P{}^{n}
 - 768 \bar P{}_{m} \bar P{}_{n} D{}_{r}\bar P{}_{s} D{}^{r}P{}^{s} P{}^{m} P{}^{n} 
\nn \\ &
+ 1536 \bar P{}^{r} \bar P{}^{s} D{}_{m}\bar P{}_{n} D{}_{r}P{}_{s} P{}^{m} P{}^{n} 
+ 4608 \bar P{}^{r} \bar P{}_{r} \bar P{}^{s} \bar P{}_{s} P{}^{m} P{}_{m} P{}^{n} P{}_{n} 
- 2304 \bar P{}_{n} \bar P{}_{r} \bar P{}^{s} \bar P{}_{s} P{}^{m} P{}_{m} P{}^{n} P{}^{r} 
\nn \\ &
+ 576 \bar P{}_{m} \bar P{}_{n} \bar P{}_{r} \bar P{}_{s} P{}^{m} P{}^{n} P{}^{r} P{}^{s} 
- 3072 \bar P{}^{n} D{}_{r}\bar P{}^{t} D{}^{r}P{}^{s} P{}^{m} R{}_{m}{}_{n}{}_{s}{}_{t} 
+ 3072 \bar P{}^{n} D{}_{n}\bar P{}^{t} D{}^{r}P{}^{s} P{}^{m} R{}_{m}{}_{r}{}_{s}{}_{t}
\nn \\ &
+ 3072 \bar P{}^{n} D{}_{n}P{}^{r} D{}^{s}\bar P{}^{t} P{}^{m} R{}_{m}{}_{s}{}_{r}{}_{t} 
- 3072 \bar P{}^{n} D{}_{m}\bar P{}^{t} D{}^{r}P{}^{s} P{}^{m} R{}_{n}{}_{r}{}_{s}{}_{t} 
- 3072 \bar P{}^{n} D{}_{m}P{}^{r} D{}^{s}\bar P{}^{t} P{}^{m} R{}_{n}{}_{s}{}_{r}{}_{t} 
\nn \\ &
- 1536 \bar P{}_{m} D{}^{n}P{}^{r} D{}^{s}\bar P{}^{t} P{}^{m} R{}_{n}{}_{s}{}_{r}{}_{t} 
+ 768 \bar P{}^{r} \bar P{}_{r} P{}^{m} P{}^{n} R{}_{m}{}^{s}{}^{t}{}^{u} R{}_{n}{}_{s}{}_{t}{}_{u} 
+ 1536 D{}^{m}P{}^{n} D{}^{r}\bar P{}^{s} R{}_{m}{}^{t}{}_{r}{}^{u} R{}_{n}{}_{u}{}_{s}{}_{t} 
\nn \\ &
+ 1536 \bar P{}^{r} \bar P{}^{s} P{}^{m} P{}^{n} R{}_{m}{}^{t}{}_{r}{}^{u} R{}_{n}{}_{u}{}_{s}{}_{t} 
- 1536 D{}_{m}\bar P{}^{r} D{}^{m}P{}^{n} R{}_{n}{}^{s}{}^{t}{}^{u} R{}_{r}{}_{s}{}_{t}{}_{u} 
+ 768 \bar P{}^{n} \bar P{}^{r} P{}^{m} P{}_{m} R{}_{n}{}^{s}{}^{t}{}^{u} R{}_{r}{}_{s}{}_{t}{}_{u} 
\nn \\ &
- 3072 \bar P{}_{m} \bar P{}^{r} P{}^{m} P{}^{n} R{}_{n}{}^{s}{}^{t}{}^{u} R{}_{r}{}_{s}{}_{t}{}_{u} 
+ 192 D{}_{m}\bar P{}_{n} D{}^{m}P{}^{n} R{}^{r}{}^{s}{}^{t}{}^{u} R{}_{r}{}_{s}{}_{t}{}_{u} 
- 192 \bar P{}^{n} \bar P{}_{n} P{}^{m} P{}_{m} R{}^{r}{}^{s}{}^{t}{}^{u} R{}_{r}{}_{s}{}_{t}{}_{u} 
\nn \\ &
+ 480 \bar P{}_{m} \bar P{}_{n} P{}^{m} P{}^{n} R{}^{r}{}^{s}{}^{t}{}^{u} R{}_{r}{}_{s}{}_{t}{}_{u} 
+ 1536 D{}^{m}P{}^{n} D{}^{r}\bar P{}^{s} R{}_{m}{}^{t}{}_{n}{}^{u} R{}_{r}{}_{t}{}_{s}{}_{u} 
+ 1536 \bar P{}^{r} \bar P{}^{s} P{}^{m} P{}^{n} R{}_{m}{}^{t}{}_{n}{}^{u} R{}_{r}{}_{t}{}_{s}{}_{u} \brs *_{10}1 \,.
\label{St8}
\ea

The analogous expansion of \eqref{12de12e12} gives instead 
\ba \label{eq:epepR4-expand}
S^\tbthree_{\e_{12}} &= \fr18 \int f_0(\tau,\bar{\tau})\,  \bls  \e_{10} \e_{10}  R^4 -192 D{}_{m}\bar P{}_{n} D{}^{m}P{}^{n} R{}^{2} + 192 D{}^{m}P{}_{m} D{}^{n}\bar P{}_{n} R{}^{2} 
+ 384 \bar P{}^{n} \bar P{}_{n} P{}^{m} P{}_{m} R{}^{2} 
\nn \\& 
- 384 \bar P{}_{m} \bar P{}_{n} P{}^{m} P{}^{n} R{}^{2} 
+ 32 \bar P{}_{m} P{}^{m} R{}^{3} - 768 D{}^{m}P{}^{n} D{}^{r}\bar P{}_{r} R R{}_{m}{}_{n} 
- 1536 \bar P{}^{r} \bar P{}_{r} P{}^{m} P{}^{n} R R{}_{m}{}_{n} 
\nn \\&
- 384 \bar P{}^{n} P{}^{m} R{}^{2} R{}_{m}{}_{n}
 + 768 D{}^{m}P{}^{n} D{}^{r}\bar P{}^{s} R R{}_{m}{}_{r}{}_{n}{}_{s} 
+ 1536 \bar P{}^{r} \bar P{}^{s} P{}^{m} P{}^{n} R R{}_{m}{}_{r}{}_{n}{}_{s} 
\nn \\& 
+ 1536 D{}_{m}\bar P{}^{r} D{}^{m}P{}^{n} R R{}_{n}{}_{r} 
- 768 D{}^{m}P{}_{m} D{}^{n}\bar P{}^{r} R R{}_{n}{}_{r} 
- 1536 \bar P{}^{n} \bar P{}^{r} P{}^{m} P{}_{m} R R{}_{n}{}_{r}
\nn \\& 
 + 3072 \bar P{}_{m} \bar P{}^{r} P{}^{m} P{}^{n} R R{}_{n}{}_{r} 
 + 1536 \bar P{}^{n} P{}^{m} R R{}_{m}{}^{r} R{}_{n}{}_{r}
 - 384 \bar P{}_{m} P{}^{m} R R{}^{n}{}^{r} R{}_{n}{}_{r} 
 \nn \\& 
 - 768 \bar P{}^{n} P{}^{m} R R{}_{m}{}^{r}{}^{s}{}^{t} R{}_{n}{}_{r}{}_{s}{}_{t} 
 + 3072 D{}^{m}P{}^{n} D{}^{r}\bar P{}^{s} R{}_{m}{}^{t} R{}_{n}{}_{r}{}_{s}{}_{t} 
 + 6144 \bar P{}^{r} \bar P{}^{s} P{}^{m} P{}^{n} R{}_{m}{}^{t} R{}_{n}{}_{r}{}_{s}{}_{t}
 \nn \\& 
 + 96 \bar P{}_{m} P{}^{m} R R{}^{n}{}^{r}{}^{s}{}^{t} R{}_{n}{}_{r}{}_{s}{}_{t} 
 - 1536 D{}^{m}P{}^{n} D{}^{r}\bar P{}^{s} R{}_{m}{}_{r} R{}_{n}{}_{s} 
 - 3072 \bar P{}^{r} \bar P{}^{s} P{}^{m} P{}^{n} R{}_{m}{}_{r} R{}_{n}{}_{s} 
  \nn \\& 
 + 1536 D{}^{m}P{}^{n} D{}^{r}\bar P{}_{r} R{}_{m}{}^{s} R{}_{n}{}_{s} 
 + 3072 \bar P{}^{r} \bar P{}_{r} P{}^{m} P{}^{n} R{}_{m}{}^{s} R{}_{n}{}_{s} 
 + 768 D{}^{m}P{}^{n} D{}^{r}\bar P{}^{s} R{}_{m}{}_{r}{}^{t}{}^{u} R{}_{n}{}_{s}{}_{t}{}_{u} 
  \nn \\& 
 + 1536 \bar P{}^{r} \bar P{}^{s} P{}^{m} P{}^{n} R{}_{m}{}_{r}{}^{t}{}^{u} R{}_{n}{}_{s}{}_{t}{}_{u} 
 - 768 D{}^{m}P{}^{n} D{}^{r}\bar P{}_{r} R{}_{m}{}^{s}{}^{t}{}^{u} R{}_{n}{}_{s}{}_{t}{}_{u} 
 - 1536 \bar P{}^{r} \bar P{}_{r} P{}^{m} P{}^{n} R{}_{m}{}^{s}{}^{t}{}^{u} R{}_{n}{}_{s}{}_{t}{}_{u}
  \nn \\&  
 + 1536 D{}^{m}P{}^{n} D{}^{r}\bar P{}^{s} R{}_{m}{}^{t}{}_{r}{}^{u} R{}_{n}{}_{t}{}_{s}{}_{u} 
 + 3072 \bar P{}^{r} \bar P{}^{s} P{}^{m} P{}^{n} R{}_{m}{}^{t}{}_{r}{}^{u} R{}_{n}{}_{t}{}_{s}{}_{u} 
 + 1536 \bar P{}^{n} P{}^{m} R R{}_{m}{}_{r}{}_{n}{}_{s} R{}^{r}{}^{s} 
  \nn \\& 
 + 1536 \bar P{}^{n} P{}^{m} R{}_{m}{}_{r}{}^{t}{}^{u} R{}_{n}{}_{s}{}_{t}{}_{u} R{}^{r}{}^{s} 
 + 3072 \bar P{}^{n} P{}^{m} R{}_{m}{}^{t}{}_{r}{}^{u} R{}_{n}{}_{t}{}_{s}{}_{u} R{}^{r}{}^{s} 
 + 1536 D{}^{m}P{}^{n} D{}^{r}\bar P{}^{s} R{}_{m}{}_{n} R{}_{r}{}_{s} 
   \nn \\& 
 + 3072 \bar P{}^{r} \bar P{}^{s} P{}^{m} P{}^{n} R{}_{m}{}_{n} R{}_{r}{}_{s} 
 - 3072 D{}_{m}\bar P{}^{r} D{}^{m}P{}^{n} R{}_{n}{}^{s} R{}_{r}{}_{s} 
 + 1536 D{}^{m}P{}_{m} D{}^{n}\bar P{}^{r} R{}_{n}{}^{s} R{}_{r}{}_{s} 
   \nn \\& 
 + 3072 \bar P{}^{n} \bar P{}^{r} P{}^{m} P{}_{m} R{}_{n}{}^{s} R{}_{r}{}_{s} 
 - 6144 \bar P{}_{m} \bar P{}^{r} P{}^{m} P{}^{n} R{}_{n}{}^{s} R{}_{r}{}_{s} 
 - 3072 \bar P{}^{n} P{}^{m} R{}_{m}{}^{r} R{}_{n}{}^{s} R{}_{r}{}_{s} 
   \nn \\& 
 + 512 \bar P{}_{m} P{}^{m} R{}^{n}{}^{r} R{}_{n}{}^{s} R{}_{r}{}_{s} 
 + 768 D{}_{m}\bar P{}_{n} D{}^{m}P{}^{n} R{}^{r}{}^{s} R{}_{r}{}_{s} 
 - 768 D{}^{m}P{}_{m} D{}^{n}\bar P{}_{n} R{}^{r}{}^{s} R{}_{r}{}_{s} 
   \nn \\& 
 - 1536 \bar P{}^{n} \bar P{}_{n} P{}^{m} P{}_{m} R{}^{r}{}^{s} R{}_{r}{}_{s} 
 + 1536 \bar P{}_{m} \bar P{}_{n} P{}^{m} P{}^{n} R{}^{r}{}^{s} R{}_{r}{}_{s} 
 + 1536 \bar P{}^{n} P{}^{m} R{}_{m}{}_{n} R{}^{r}{}^{s} R{}_{r}{}_{s} 
   \nn \\& 
 + 1536 \bar P{}^{n} P{}^{m} R{}_{m}{}^{s}{}^{t}{}^{u} R{}_{n}{}^{r} R{}_{r}{}_{s}{}_{t}{}_{u} 
 + 1536 D{}_{m}\bar P{}^{r} D{}^{m}P{}^{n} R{}_{n}{}^{s}{}^{t}{}^{u} R{}_{r}{}_{s}{}_{t}{}_{u} 
 - 768 D{}^{m}P{}_{m} D{}^{n}\bar P{}^{r} R{}_{n}{}^{s}{}^{t}{}^{u} R{}_{r}{}_{s}{}_{t}{}_{u} 
   \nn \\& 
 - 1536 \bar P{}^{n} \bar P{}^{r} P{}^{m} P{}_{m} R{}_{n}{}^{s}{}^{t}{}^{u} R{}_{r}{}_{s}{}_{t}{}_{u} + 3072 \bar P{}_{m} \bar P{}^{r} P{}^{m} P{}^{n} R{}_{n}{}^{s}{}^{t}{}^{u} R{}_{r}{}_{s}{}_{t}{}_{u} + 1536 \bar P{}^{n} P{}^{m} R{}_{m}{}^{r} R{}_{n}{}^{s}{}^{t}{}^{u} R{}_{r}{}_{s}{}_{t}{}_{u} 
   \nn \\& 
 - 768 \bar P{}_{m} P{}^{m} R{}^{n}{}^{r} R{}_{n}{}^{s}{}^{t}{}^{u} R{}_{r}{}_{s}{}_{t}{}_{u} 
 - 192 D{}_{m}\bar P{}_{n} D{}^{m}P{}^{n} R{}^{r}{}^{s}{}^{t}{}^{u} R{}_{r}{}_{s}{}_{t}{}_{u} 
 + 192 D{}^{m}P{}_{m} D{}^{n}\bar P{}_{n} R{}^{r}{}^{s}{}^{t}{}^{u} R{}_{r}{}_{s}{}_{t}{}_{u} 
   \nn \\& 
 + 384 \bar P{}^{n} \bar P{}_{n} P{}^{m} P{}_{m} R{}^{r}{}^{s}{}^{t}{}^{u} R{}_{r}{}_{s}{}_{t}{}_{u} 
 - 384 \bar P{}_{m} \bar P{}_{n} P{}^{m} P{}^{n} R{}^{r}{}^{s}{}^{t}{}^{u} R{}_{r}{}_{s}{}_{t}{}_{u} 
 - 384 \bar P{}^{n} P{}^{m} R{}_{m}{}_{n} R{}^{r}{}^{s}{}^{t}{}^{u} R{}_{r}{}_{s}{}_{t}{}_{u}
   \nn \\& 
  - 3072 D{}^{m}P{}^{n} D{}^{r}\bar P{}^{s} R{}_{m}{}_{s}{}_{n}{}_{t} R{}_{r}{}^{t} 
  - 6144 \bar P{}^{r} \bar P{}^{s} P{}^{m} P{}^{n} R{}_{m}{}_{s}{}_{n}{}_{t} R{}_{r}{}^{t} 
  - 3072 \bar P{}^{n} P{}^{m} R{}_{m}{}_{s}{}_{n}{}_{t} R{}^{r}{}^{s} R{}_{r}{}^{t} 
    \nn \\& 
  - 1536 D{}^{m}P{}^{n} D{}^{r}\bar P{}^{s} R{}_{m}{}^{t}{}_{n}{}^{u} R{}_{r}{}_{t}{}_{s}{}_{u}
   - 3072 \bar P{}^{r} \bar P{}^{s} P{}^{m} P{}^{n} R{}_{m}{}^{t}{}_{n}{}^{u} R{}_{r}{}_{t}{}_{s}{}_{u} 
   - 3072 \bar P{}^{n} P{}^{m} R{}_{m}{}^{t}{}_{n}{}^{u} R{}^{r}{}^{s} R{}_{r}{}_{t}{}_{s}{}_{u} 
     \nn \\& 
   + 3072 \bar P{}^{n} P{}^{m} R{}_{m}{}^{r}{}^{s}{}^{t} R{}_{n}{}^{u}{}_{s}{}^{v} R{}_{r}{}_{v}{}_{t}{}_{u}
    - 256 \bar P{}_{m} P{}^{m} R{}^{n}{}^{r}{}^{s}{}^{t} R{}_{n}{}^{u}{}_{s}{}^{v} R{}_{r}{}_{v}{}_{t}{}_{u} 
    + 1536 D{}^{m}P{}^{n} D{}^{r}\bar P{}_{r} R{}_{m}{}_{s}{}_{n}{}_{t} R{}^{s}{}^{t} 
      \nn \\& 
    + 3072 \bar P{}^{r} \bar P{}_{r} P{}^{m} P{}^{n} R{}_{m}{}_{s}{}_{n}{}_{t} R{}^{s}{}^{t}
    - 3072 \bar P{}^{n} P{}^{m} R{}_{m}{}_{s}{}_{r}{}_{t} R{}_{n}{}^{r} R{}^{s}{}^{t} 
    - 3072 D{}_{m}\bar P{}^{r} D{}^{m}P{}^{n} R{}_{n}{}_{s}{}_{r}{}_{t} R{}^{s}{}^{t}
      \nn \\&  
    + 1536 D{}^{m}P{}_{m} D{}^{n}\bar P{}^{r} R{}_{n}{}_{s}{}_{r}{}_{t} R{}^{s}{}^{t} 
    + 3072 \bar P{}^{n} \bar P{}^{r} P{}^{m} P{}_{m} R{}_{n}{}_{s}{}_{r}{}_{t} R{}^{s}{}^{t}
     - 6144 \bar P{}_{m} \bar P{}^{r} P{}^{m} P{}^{n} R{}_{n}{}_{s}{}_{r}{}_{t} R{}^{s}{}^{t} 
       \nn \\& 
      - 3072 \bar P{}^{n} P{}^{m} R{}_{m}{}^{r} R{}_{n}{}_{s}{}_{r}{}_{t} R{}^{s}{}^{t} 
      + 768 \bar P{}_{m} P{}^{m} R{}^{n}{}^{r} R{}_{n}{}_{s}{}_{r}{}_{t} R{}^{s}{}^{t} 
      - 768 \bar P{}^{n} P{}^{m} R{}_{m}{}^{r}{}^{s}{}^{t} R{}_{n}{}_{r}{}^{u}{}^{v} R{}_{s}{}_{t}{}_{u}{}_{v} 
        \nn \\& 
      + 64 \bar P{}_{m} P{}^{m} R{}^{n}{}^{r}{}^{s}{}^{t} R{}_{n}{}_{r}{}^{u}{}^{v} R{}_{s}{}_{t}{}_{u}{}_{v} + 1536 \bar P{}^{n} P{}^{m} R{}_{m}{}^{r}{}_{n}{}^{s} R{}_{r}{}^{t}{}^{u}{}^{v} R{}_{s}{}_{t}{}_{u}{}_{v} \brs *_{10} 1 \,.
\ea

\end{appendix}



\end{document}